\documentclass[a4paper]{PoS}

\DeclareMathAlphabet{\mathcal}{OMS}{cmsy}{m}{n}

\newcommand{\etal}{{et al.}}

\usepackage{lineno}

\title{A high-level analysis framework for HAWC}

\ShortTitle{HAWC analysis framework}

\author{Patrick W. Younk $^{a}$ , \speaker{Robert J. Lauer} $^{b}$, Giacomo Vianello $^{c}$, J. Patrick Harding $^{a}$, Hugo Alberto Ayala Solares $^{d}$,
Hao Zhou $^{d}$, Michelle Hui $^{d}$, for the HAWC Collaboration $^{e}$\\
        \llap{$^a$}
        Los Alamos National Laboratory, Physics Division, Los Alamos, NM, USA\\
        \llap{$^b$}
        University of New Mexcio, Department of Physics and Astronomy, Albuquerque, NM, USA\\
        \llap{$^c$}
        Stanford University, Department of Physics, Stanford, CA, USA\\
        \llap{$^d$}
        Michigan Technological University, Department of Physics, Houghton, MI, USA\\
        \llap{$^e$}
        For a complete author list, see \href{http://www.hawc-observatory.org/collaboration/icrc2015.php}{www.hawc-observatory.org/collaboration/icrc2015.php}\\
        E-mail: \email{rlauer@phys.unm.edu}
        }

\abstract{
The High Altitude Water Cherenkov (HAWC) Observatory continuously observes gamma-rays between 100 GeV to 100 TeV in an instantaneous field of view of about 2 steradians above the array.
The large amount of raw data, the importance of small number statistics, the large dynamic range of gamma-ray signals in time (1 - $10^8$ sec) and angular extent (0.1 - 100 degrees), and the growing need to directly compare results from different observatories pose some special challenges for the analysis of HAWC data. 
To address these needs, we have designed and implemented a modular analysis framework based on the method of maximum likelihood. The framework facilitates the calculation of a binned Poisson Log-likelihood value for a given physics model (i.e., source model), data set, and detector response. 
The parameters of the physics model (sky position, spectrum, angular extent, etc.) can be optimized through a likelihood maximization routine to obtain a best match to the data. 
In a similar way, the parameters of the detector response (absolute pointing, angular resolution, etc.) can be optimized using a well-known source such as the Crab Nebula.
The framework was designed concurrently with the Multi-Mission Maximum Likelihood (3ML) architecture, and allows for the definition of a general collection of sources with individually varying spectral and spatial morphologies. Compatibility with the 3ML architecture allows to easily perform powerful joint fits with other observatories. In this contribution, we overview the design and capabilities of the HAWC analysis framework, stressing the overarching design points that have applicability to other astronomical and cosmic-ray observatories.}

\FullConference{The 34th International Cosmic Ray Conference,\\
		30 July- 6 August, 2015\\
		The Hague, The Netherlands}

\begin{document}


\section{Introduction}

The High Altitude Water Cherenkov (HAWC) Gamma-ray Observatory is a newly commissioned instrument on volcano Sierra Negra in the state of Puebla, Mexico~\cite{bib:hawcsensi}.
Here we describe the design and capabilities of a high-level (i.e., event-level) analysis framework.
The framework is a set of software objects that
provides particular functionality for statistical analysis of event-level data and that facilitates the development of analysis programs by end users.
Event-level data are air shower events observed in the HAWC detector that have been reconstructed (i.e., a direction
and other shower observables have been estimated) 
and have passed some low-level selection triggers. 
Our framework facilitates statistical analysis using the well-known likelihood formalism \cite{Neyman}.
The framework supports a model-centric approach for a user to design an analysis program.
That is, a user of the framework mainly concentrates on the physics model or detector model that she wants to test.
The software interfaces to the physics model, the detector response, the binned data, and the calculation of likelihood values are standardized.
The actual details of the physics models, the details of the detector response, and the binning of the data are defined by the user through these standardized interfaces.

Some noteworthy design features of the framework are:
\begin{itemize}
  \item The physics model definition has no restriction with regard to energy spectrum, source spatial morphology, or number of sources.
  \item The physics model can be arbitrarily parameterized and any number of parameters can be optimized with a maximum likelihood routine.
  \item The detector response model can be parameterized enabling details of the detector response to be estimated directly from data 
using a well-known gamma-ray source (e.g., the Crab Nebula).
  \item The data is binned on the sky using Healpix bins~\cite{bib:healpix}. The data can be binned in an arbitrary number of other dimensions (e.g., event quality, energy, etc.) that are defined by the user. The binned data (the size of which is typically $~10^8$ bins, neglecting the time dimension) is efficiently stored on disk and handled in memory.
  \item A routine to calculate a profile maximum likelihood is supported. 
  \item The framework is fully compatible with the Multi-mission Maximum Likelihood (3ML) architecture \cite{Vianello_icrc} which enables the
incorporation of multiple instruments (e.g., HAWC and Fermi/LAT) in a joint-likelihood fit.
\end{itemize}

In the following sections we elucidate the design features of the 
HAWC event-level analysis framework, also called \textbf{Li}kelihood \textbf{F}itting \textbf{F}ramework (LiFF).


\section{Likelihood Formalism for Statistical Significance}

The main purpose of the framework  is to facilitate the calculation of the likelihood of a particular physics model and detector response model given an observed set of data.
Currently, the framework is designed with the minimum time window as one transit (one day)~\footnote{We currently have other software routines for dealing with shorter
time windows. We plan to expand the capabilities of this framework to shorter time windows in the future.}. 
To substantially reduce the computational workload, we bin the data -- we calculate a binned likelihood value.
%
HAWC event-level data is binned in a multi-dimensional array of event attributes. For a typical HAWC analysis, the array dimensions include the event direction 
as a set of two sky coordinates (e.g., right ascension and declination)
and event quality and/or an energy metric.

The calculation of the logarithm of the likelihood is computationally convenient:
$$ \ln \mathcal{L} (\theta; \mathbf{N}_{obs}) = \sum_{i = 0}^{All Bins} \ln(f((N_{obs})_i | \theta )),$$
where $\theta$ is a set of model parameters, $(N_{obs})_i$ is the number of events in the ith bin, $\mathbf{N}_{obs}$ is the collection of all $(N_{obs})_i$,
and $f$ is the probability of observing $(N_{obs})_i$ given $\theta$.
To estimate the optimum set of parameters $\theta_0$, we find the $\theta$ that maximizes the value of $\ln \mathcal{L}$,
or equivalently we minimize the value of $-\ln \mathcal{L}$.
Estimates of the parameter uncertainties and the correlations between parameters can be calculated with well known methods \cite{UncertaintyEstimation}.

To compare hypotheses, the Likelihood Ratio test is used. A test statistic is defined as
$$ TS = 2 \ln \frac{ \mathcal{L} (\mbox{Alternative Hypothesis} ; \mathbf{N}_{obs})}  { \mathcal{L} (\mbox{Null Hypothesis}; \mathbf{N}_{obs})}.$$
When the null-hypothesis is true and in the case of nested models, $TS$ is approximately distributed as the $\chi^2$ distribution
with the number of degrees of freedom equal to the difference in the number of free parameters between the hypotheses (Wilks' theorem \cite{bib:Wilks}).
For example, the usual source discovery test has the null hypotheses as ``there is no source'' and the alternative hypothesis as ``there is a source with a flux normalization $X$''.
In this case, the difference between the number of free parameters is one (the flux normalization). When the null hypothesis is true, $TS$ is distributed as $\chi^2(DOF = 1)$, 
and $\sqrt{TS}$ can be interpreted as the significance of the over-abundance of events in units of "Gaussian sigma". 
Using the $TS$ formalism enables comparison of arbitrary hypotheses 
with the full rigor of modern frequentist statistics. Note that non-nested models cannot use the Wilks' theorem approximation but there
are other methods available \cite{Non-nested}.

When the expected number of signal events in a bin is small (e.g., either due to a weak source, or a short observation time, or both), 
Poisson probability functions must be used in the calculation of $\ln \mathcal{L}$ or the calculation
of the $TS$ will be biased and not easily interpreted. 
In general, when using Poisson probability functions, $\ln \mathcal{L}$ is nonlinear with regard to the model parameters $\theta$.
Therefore, the minimization of the log-likelihood function is performed with a iterative, gradient search algorithm. We use the Minuit \cite{Minuit} minimization package.
If we assume a Gaussian probability function instead of Poisson, $\ln \mathcal{L}$ is linear with regard to the flux normalization parameter and can be solved algebraically.
This method is known as Gaussian weighting in the literature (e.g., \cite{GaussianWeighting}). Our framework incorporates a method that implements 
the Gaussian weighting algorithm. The results of the Gaussian weighting algorithm are typically used as 
a first guess for the minimization of the non-linear $\ln \mathcal{L}$ function.

The model parameters $\theta$ can be divided into physics parameters (the parameters of main interest) and nuisance parameters.
An example of a nuisance parameter for HAWC is the background normalization (the amount of hadronic events that pass event quality cuts) in a particular bin.
A profile maximum likelihood takes into account the idea of physics parameters and nuisance parameters.
Two nested loops are implemented where the nuisance parameters are estimated in the inner loop and the physics parameters
are estimated in the outer loop. 
In this way, the outer loop can incorporate likelihood values for a particular physics model from several different instruments with the nuisance parameters
associated with each instrument effectively decoupled. This idea is part of the 3ML architecture, which is described in detail in \cite{Vianello_icrc}.
The 3ML idea can be thought of as standardizing the statistical language that is used both within a collaboration
and between collaborations -- it has wide applicability.


\section{Software Architecture}

In Figure 1 we show a diagram of the framework architecture and how it is used in the analysis of HAWC data.
The framework is composed of software objects grouped into five modules: 1.) Physics Model, 2.) Detector Response Model, 3.) Binned Data, 4.) Likelihood Calculation, and 5.) Minimizer.
The dependencies of the framework are \textsc{Root} \cite{Root} and HEALPix.
In the following sub-sections we briefly overview the highlights of the framework components. 

\begin{figure}
\includegraphics[width=1.0\textwidth]{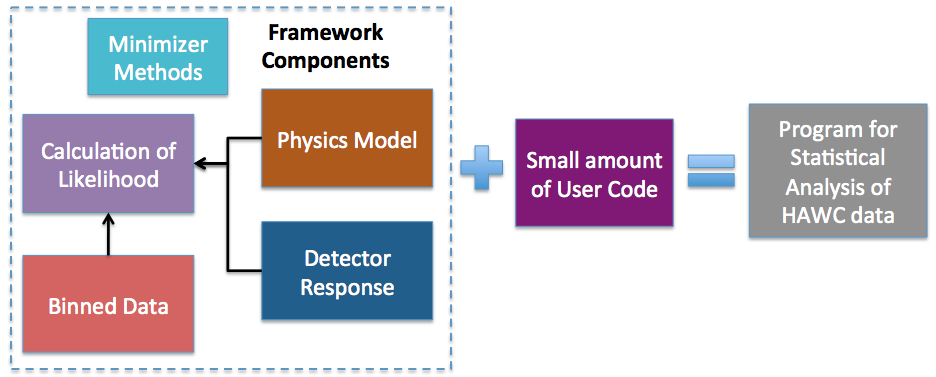}
\caption{A diagram of the framework architecture. The code writting by the user is typically not complex. For example, software objects are instantiated and initialized, 
then a \textit{GetLikelihood} method is called.}
\label{fig1}
\end{figure}


\subsection{Binned Event-level Data}

Event-level data is binned on the sky using HEALPix pixels, which are equal area pixels. A set of HEALPix pixels that covers the whole sky is called a HEALPix Map.
Event-level data are also binned in other dimensions. 
For example, we currently typically use 10 event-quality bins. The events in a particular event-quality bin have a similar angular resolution and gamma-hadron resolution.
That is, the detector response is a function of the event-quality. Our current metric for event-quality is the number of hit photomultiplier tubes in the array,
which is strongly correlated to the energy deposited in the array.
One HEALPix map with event counts is stored for each event-quality bin.

An arbitrary number of bin dimensions beyond the two sky coordinates can be defined and implemented by the user.
Each additional bin, in a dimension other than the sky coordinates, is realized as an additional HEALPix Map. 
For example, the data can be binned in both event-quality and in a metric which estimates the energy of the primary particle.
This energy metric is the energy deposited on the array adjusted for core location on the array and zenith angle.
Binning in an energy metric is important for detailed spectral analysis.
If we had 10 event quality bins and 10 energy bins, we would have a total of 100 HEALPix Maps. In this way, a HEALPix map is a "higher dimension bin". 



In a typical analysis using our framework, only the bins that have a signal expectation according to the Physics Model and Detector Response Model
are summed over in the calculation of $\ln \mathcal{L}$.
For a Physics Model that is a point-source, for example, only sky pixels in a region of typically 5$^{\circ}$ radius around the hypothesized source location
have to be analyzed, since the point spread function of HAWC vanishes for larger angular separations.
Thus, we typically only keep a small fraction of the total number of sky-pixels in memory at any one time.
This significantly lowers our memory requirements (important if jobs are to be run in parallel) as well as our disk access time.
To handle partial sky regions in memory, we developed the \textit{SkyMap} class that inherits from the HEALPix base class.
The \textit{SkyMap} class stores a limited number of sky pixels where the boundaries of the pixel-set can be arbitrary polygonal or circular shapes.

As a fast and compressed file storage format for HEALPix maps, we implemented a class called \textit{MapTree} that depends on \textsc{Root} object classes. 
Each pixel value is stored in a \textit{TTree} object with a single branch, 
where the entry index number is identical to the HEALPix pixel index, and the \textit{TTree} is stored as a \textsc{Root} \textit{TFile}. 
We have found that with this on-disk format, writing and reading times for a full set of sky pixels is faster than or comparable to 
the FITS~\cite{bib:fits} on-disk format. More importantly, the writing and reading of partial sky regions are substantially faster.
%

Our time bin, the length of time for which data is collected to generate a single set of HEALPix maps, is an arbitrary time period.
The maximum length of the time bin is typically determined by the duration of a stable detector configuration, relevant in particular during the construction phase. To facilitate time-dependent analyzes like in~\cite{bib:agn}, the current minimum time interval is one sidereal day.


\subsection{Physics Model}

The physics model is implemented as a collection of one or more sources. We define two types of sources: 1.) A point source, and 2.) An extended source. 
The point source has a location and a spectrum. 
A general spectrum is defined by a table of differential flux values, binned in arbitrary energy intervals. For many analyzes, the spectrum can also be described by a simple function, for example a power law, in which case the spectrum is stored as a \textit{TF1} object \cite{Root}.
The extended source has a location, a spectrum, and a spatial morphology. 
The spatial morphology interface is still under development but will eventually allow arbitrary tabular or functional inputs describing flux intensity and spectra as a function of location.

This implementation gives the user a tremendous amount of freedom in setting the physics model. 
Note that each source can have an independent spectrum and that point sources and extended sources can be implemented in the same physics model.
Some examples of physics models that can be implemented and tested are
a closely spaced double source, a soft-spectrum source embedded in a hard-spectrum extended source, and the diffuse emission from the galactic plane.


\subsection{Detector Response Model}

If the time bin is an integer number of sidereal days, then the detector response is a function of the declination of the source but not the right ascension.
We define the detector response as a function of the declination and the source spectrum.
Given a source declination and arbitrary source spectrum 
the code within the Detector Response Module calculates the expected point spread functions
(how events are spread on the sky due to detection limitations) 
for each of the higher-dimension bins and the number of events in each of the higher-dimension bins.
This information is then used by code within the Likelihood Calculation Module to determine the expected number of signal events in each bin for a particular
physics model.

The description of the detector response is realized as a collection of histograms (\textit{TH1} objects
~\cite{Root}). These histograms are generated from simulation using
a particular spectrum. If the source spectrum called for in the physics model is different from the spectrum used to generate the original detector response, the detector
response is modified by re-weighting the energy distribution histograms for each higher dimension bin. This is a substantial computational short cut and is accurate as long
as the spectrum called for in the physics model is not too different from the spectrum used to generate the original detector response.

There is functionality such that the detector response can be parameterized and these parameters can be free in the minimization of $- \ln \mathcal{L}$.
This allows for the determination of the detector response directly from the data.
For example, if we assume that the spatial extent of the Crab Nebula is much smaller than our point spread function, 
we can determine the point spread function that is most consistent with the collection of events with directions near the Crab~\cite{bib:icrccrab}.
%


\subsection{Likelihood Calculation}

To calculate $\ln \mathcal{L}$ we need to know the expected number of signal events and the expected number of background events for each bin.
%
%
The expected number of signal events in each relevant bin (bins that have zero expected source events are not included) is calculated 
using the physics model parameters set by the user and the relevant detector response information.
The expected number of background events (i.e., hadron events that have passed our event quality selections) are normally not estimated from simulation but rather directly from the data. 
The background is either estimated in a prior step that occurs when the data is first binned \cite{Andy'sContribution}, or the
background is estimated as a nuisance parameter as part of the minimization of $-\ln \mathcal{L}$.

Then to calculate the log likelihood using Poisson probability:
$$ \ln \mathcal{L} = \sum^{All Bins} (N_{obs} \ln N_{exp} - N_{exp} - \ln(N_{obs} !)).$$
Note that calculating $(\ln (x!))$ is numerically efficient (e.g., using Stirling's approximation).

A minimization of $-\ln \mathcal{L}$ can be performed with an arbitrary number of free parameters.
Two nested minimization loops are supported. This enables the calculation of a profile likelihood where nuisance parameters (e.g., the
expected number of background events in each bin) are estimated in the inner loop and the parameters associated with the physics model
are estimated in the outer loop. The calculation of such a profile likelihood is key to incorporating multiple instruments (e.g., HAWC and Fermi/LAT) in a joint-Likelihood fit.

\section{Conclusions}

The HAWC Collaboration has designed and coded a software framework for the statistical analysis of event-level data. The likelihood formalism is used.
With the addition of a small amount of code, a user can efficiently explore any number of complex hypotheses.
The framework is model-centric in that the user mainly focuses on the design of the physical model that she wants to confront with data.
The complexities of handling the binned data, the detector response, the calculation of $\ln \mathcal{L}$, and the minimizations are handled by
standardized software interfaces.

The framework was designed along side and is fully compatible with the 3ML architecture. 
The 3ML architecture allows for the incorporation of multiple instruments in a joint-likelihood fit and is a powerful addition to the broader field
of astronomy and cosmic-ray physics.

%

\section*{Acknowledgments}
\footnotesize{
We acknowledge the support from: the US National Science Foundation (NSF);
the US Department of Energy Office of High-Energy Physics;
the Laboratory Directed Research and Development (LDRD) program of
Los Alamos National Laboratory; Consejo Nacional de Ciencia y Tecnolog\'{\i}a (CONACyT),
Mexico (grants 260378, 55155, 105666, 122331, 132197, 167281);
Red de F\'{\i}sica de Altas Energ\'{\i}as, Mexico;
DGAPA-UNAM (grants IG100414-3, IN108713,  IN121309, IN115409, IN113612);
VIEP-BUAP (grant 161-EXC-2011);
the University of Wisconsin Alumni Research Foundation;
the Institute of Geophysics, Planetary Physics, and Signatures at Los Alamos National Laboratory;
the Luc Binette Foundation UNAM Postdoctoral Fellowship program.
}

\end{document}